\documentclass[a4paper,11pt]{article}
\pdfoutput=1 
\usepackage{soul} 
\usepackage{jcappub} 
\usepackage[T1]{fontenc} 
\usepackage{graphicx}
\usepackage{bm}
\newcommand{\be}{\begin{eqnarray}}
\newcommand{\ee}{\end{eqnarray}}
\newcommand{\Lag}{{\cal L}}
\newcommand{\barp}{{\bar\phi}}

\subheader{ICCUB-15-035}
\title{On post-inflation validity of perturbation theory in Horndeski scalar-tensor models}

\author[a]{Cristiano Germani,}
\author[b]{Nina Kudryashova,}
\author[c]{Yuki Watanabe}

\affiliation[a]{Institut de Ci\`encies del Cosmos (ICCUB), Universitat de Barcelona, Mart\'i Franqu\`es 1, E08028 Barcelona, Spain}
\affiliation[b]{Arnold Sommerfeld Center, Ludwig-Maximilians-University,
Theresienstr. 37, 80333 Muenchen, Germany}
\affiliation[c]{Department of Physics, National Institute of Technology, Gunma College, Gunma 371-8530, Japan}

\emailAdd{germani@icc.ub.edu}
\emailAdd{nina.kudryashova@campus.lmu.de}
\emailAdd{yuki.watanabe@nat.gunma-ct.ac.jp}

\abstract{By using the newtonian gauge, we re-confirm that, as in the minimal case, the re-scaled Mukhanov-Sasaki variable is conserved leading to a constraint equation for the Newtonian potential. However, conversely to the minimal case, in Horndeski theories, the super-horizon Newtonian potential can potentially grow to very large values after inflation exit. If that happens, inflationary predictability is lost during the oscillating period. When this does not happen, the perturbations generated during inflation can be standardly related to the CMB, if the theory chosen is minimal at low energies. As a concrete example, we analytically and numerically discuss the new Higgs inflationary case. There, the Inflaton is the Higgs boson that is non-minimally kinetically coupled to gravity. During the high-energy part of the post-inflationary oscillations, the system is anisotropic and the Newtonian potential is largely amplified. Thanks to the smallness of today's amplitude of curvature perturbations, however, the system stays in the linear regime, so that inflationary predictions are not lost. At low energies, when the system relaxes to the minimal case, the anisotropies disappear and the Newtonian potential converges to a constant value. We show that the constant value to which the Newtonian potential converges is related to the frozen part of curvature perturbations during inflation, precisely like in the minimal case.}

\begin{document}
\maketitle
\flushbottom
\section{Introduction}
One of the most striking features of inflation is that tiny primordial quantum fluctuations of the universe spatial curvature (we will define this precisely later on) are amplified to macroscopic values \cite{Chibisov:1982nx, Mukhanov:2005sc}.
Because of the universe evolution, the wavelengths of these fluctuations eventually stretch over the cosmological horizon. At this point, curvature fluctuations freeze to a constant value, at least in "standard" minimally coupled single-field inflation.\footnote{Of course, strictly speaking the constant solution is only achieved for an infinitely long wavelength. By "standard" we mean that gravity is described by General Relativity (GR) throughout this paper.}

After the universe is reheated, during the radiation period, the imprint of inflation in terms of the curvature fluctuations is finally deposited into the cosmic microwave background radiation (CMB). The initial temperature contrast of the CMB is indeed proportional to the amplitude of the Newtonian potential ($\Psi$) that re-entered the horizon during the radiation era. The relation between the Newtonian potential and the curvature perturbations is straightforward in any generic theory of inflation: for any given equation of state of the effective fluid driving the universe, the Newtonian potential approximately settles to an order one constant times the amplitude of the curvature perturbations \cite{Dodelson:2003ft}. Thus the CMB directly brings information about inflation.

In addition to scalar perturbations, inflation generically generates a spectrum of primordial gravitational waves \cite{Starobinsky:1979ty}. Those are also frozen on super-horizon scales and start to evolve again as soon as they re-enter the horizon. The measure of these primordial gravitational waves would be a strong evidence for inflation. Unfortunately though, there is no evidence of tensor modes in the sky yet, thus, at the moment we can only state that inflation is compatible to the CMB. Because of null results, the tensor-to-scalar perturbations $r$ is severely constrained by the joint data Planck-BICEP2-Keck array VI to be smaller than $0.09$ at $95\%$ C.L. \cite{Array:2015xqh}.

All predictions of inflation discussed so far are related to the fact that, in standard single-field inflation, curvature perturbations are conserved on super-horizon scales. In \cite{Weinberg:2003sw} it has been argued that the conservation of curvature perturbations on super-horizon scales can only be achieved under adiabaticity. In other words, Weinberg has pointed out that
for an infinitely long wavelength, there is a conserved quantity (related to curvature perturbations) if and only if
\be\label{ad}
\Delta\equiv\frac{\dot\rho\delta p-\dot p\delta\rho}{3(\rho+p)^2}
\ee
vanishes. In \eqref{ad} $\rho$ and $p$ are respectively the (effective) energy density and (effective) pressure driving the Universe evolution. Finally, $\delta$ refers to the linear perturbations.

The Weinberg theorem, however, assumes a constant effective Planck mass \footnote{We thank Filippo Vernizzi for pointing this out to us.}. Suppose instead that the system we would like to consider has a time-varying Planck scale $M_*$. In this case, the gravity equations will be $G_{\alpha\beta}=\frac{T_{\alpha\beta}}{M_*^2}$. By re-defining $\rho=\rho_* \frac{M_*^2}{M_p^2}$ and $p=p_* \frac{M_*^2}{M_p^2}$, we would then get the "Einstein" frame equations $G_{\alpha\beta}=\frac{T_{\alpha\beta}^*}{M_p^2}$. If $\rho_*$ and $p_*$ are adiabatic, it follows, from the Weinberg theorem, that there exists a conserved quantity, although $\Delta\propto\dot M_*\neq 0$ (in other words $\Delta^*=0$). This is precisely what we are going to find.\footnote{Ref.~\cite{gao} is an explicit confirmation of the Weinberg theorem and \cite{langlois} that there is a conserved quantity in Horndeski theories.} 

One may then wonder why to declare $\rho$, instead of $\rho_*$, as the effective energy density. The reason is that, as we are going to assume that the standard model is minimally coupled to gravity, gravitational waves will "see" a time varying Planck mass $M_*$ and thus the physical energy density and pressure will indeed be $\rho$ and $p$, as used in the literature (see e.g.\cite{Gleyzes:2014rba}). In this sense then, although matter is non-adiabatic, the system still have a conserved super-horizon quantity.

Consider now a general scalar field non-minimally coupled to gravity. All possible non-minimal interactions of the scalar with gravity that are still second order were classified by Horndeski~\cite{Horndeski:1974wa} and are given by the Lagrangian densities~\cite{Deffayet:2011gz,Kobayashi:2011nu}
\be
\Lag_2&=&K(\phi,X)\ ,\cr
\Lag_3&=&-G_3(\phi,X)\square\phi\ ,\cr
\Lag_4&=&G_4(\phi,X)R+G_{4X}(\phi,X)\left[(\square\phi)^2-\phi_{;\mu\nu}\phi^{;\mu\nu}\right]\ ,\cr
\Lag_5&=&G_5(\phi,X)G_{\mu\nu}\phi^{;\mu\nu}-\frac{1}{6}G_{5X}(\phi,X)\left[(\square\phi)^3+2\phi_{;\mu}{}^\nu\phi_{;\nu}{}^\alpha\phi_{;\alpha}{}^\mu-3\phi_{;\mu\nu}\phi^{;\mu\nu}\square\phi\right]\ ,
\ee
where $X\equiv-\partial_\mu\phi\partial^\mu\phi/2$ and the indices $X$ and $\phi$ are a short notation for $\partial_X$ and $\partial_\phi$.

Expanding around a flat Friedmann-Lema\^itre-Robertson-Walker (FLRW) background $ds^2=-dt^2+a^2(t) d{\bm x}\cdot d{\bm x}$, and a time dependent scalar $\phi(t,{\bm x})=\bar\phi(t)+\delta\phi(t,{\bm x})$, by using the results of \cite{Bellini:2014fua} and \cite{Gleyzes:2014rba} one has in the Newtonian gauge
\be\label{delta}
\Delta\propto \frac{\dot M_*}{M_*}\frac{w}{(1+w)}H\frac{\delta\phi}{\dot{\bar\phi}}\ .
\ee
In \eqref{delta} $w\equiv p/\rho$ where the effective pressure and density are \cite{Gleyzes:2014rba}
\be
\rho \equiv T^t{}_t&=& -K+2X(K_X-G_{3\phi})+6\dot\phi H(XG_{3X}-G_{4\phi}-2XG_{4\phi X})\nonumber  \\
&&+12H^2X(G_{4X}+2XG_{4XX}-G_{5\phi}-XG_{5\phi X})+4\dot\phi H^3X(G_{5X}+XG_{5XX}),\nonumber \\
p \equiv \frac13 T^i{}_i &=& K-2X(G_{3\phi}-2G_{4\phi\phi})+4\dot\phi H(G_{4\phi}-2XG_{4\phi X}+XG_{5\phi\phi})\nonumber \\
&&-M_*^2\alpha_BH\frac{\ddot\phi}{\dot\phi}-4H^2X^2G_{5\phi X}+2\dot\phi H^3XG_{5X} .
\ee
Here $M_*$ is instead the effective Planck mass defined as \cite{Bellini:2014fua}
\be
M_*^2=2(G_4-2XG_{4X}+XG_{5\phi}-\dot\phi H XG_{5X})\ .
\ee
We see then that, generically, switching on $G_4$ and/or $G_5$ would make $\Delta\neq 0$ unless $w=0$.\footnote{For example this happens for $K=G_3=0$, $G_4=const$ and $G_5=-\phi/(2 M^2)$.} However, although this implies that the variables $\rho$ and $p$ are non-adiabatic, as already discussed in general terms, their non-adiabaticity is solely due to an effective time varying Planck constant. Thus, by rescaling pressure and density by $M_*$ one can immediately conclude that the system {\it is} adiabatic.

\section{The minimally coupled case}

All this section is a fast review of very standard results, one can find them in \cite{Mukhanov:2005sc} but also a nicely written reference paper may be \cite{Finelli:1998bu}. See also \cite{kodama, nambu} for earlier discussions of cosmological perturbations in a phase dominated by an oscillatory scalar field.

Let us consider scalar perturbations on a flat FLRW spacetime. In Newtonian gauge, the perturbed metric is \cite{Mukhanov:2005sc} (we focus only on scalar perturbations)
\be
ds^2=-(1+2\Psi)dt^2+a^2(t)(1-2\Phi)dx^idx^j\delta_{ij}\ ,
\ee
where $\Psi$ is the Newtonian potential. Consider now a minimally coupled scalar field with action
\be
A=\int d^4x \sqrt{-g}\left[-\frac{1}{2}\partial_\mu\phi\partial^\mu\phi-V(\phi) +\frac12 M_p^2R\right]\ .
\ee
The spatial traceless part of the Einstein equations immediately implies the absence of anisotropies, i.e. $\Psi=\Phi$. Expanding the scalar field into background and perturbations, i.e. $\phi=\barp+\delta\phi$, we seemingly have two scalar degrees of freedom propagating, $\Phi$ and $\delta\phi$. Of course, that is just a gauge artefact, and only one scalar degree of freedom propagates. Indeed, $\Phi$ and $\delta\phi$ are related in the Einstein equation via the momentum constraint equation. One may then decide to work with either of the two, or with a combination of them. It turns out that the gauge invariant combination, called the Mukhanov-Sasaki variable,~\cite{Mukhanov:1988, Sasaki:1986}
\be\label{conservation}
Q=\delta\phi+\frac{\dot{\bar\phi}}{H}\Phi\ ,
\ee
is especially smart. First of all, it turns out that once $\delta\phi$ is solved in favour of $\Phi$, the coefficients of the differential equations governing the Newtonian potential evolution diverge during the oscillating period proportionally to $\dot\barp^{-1}$. On the contrary, the canonical form of the differential equation governing the evolution of $Q$ ($\ddot Q+\ldots=0$) contains only finite coefficients. More importantly, for wavelengths larger than the Hubble horizon, in which spatial derivatives can be neglected (super-horizon modes), one finds an extremely simple conservation equation
\be
\partial_t\left(\frac{H}{\dot\barp}Q\right)=0\ .
\ee
Whenever the velocity of the scalar field is sign-definite (future-directed), one can define the physical curvature perturbations on a uniform density slicing. In this case, the rescaled Mukhanov-Sasaki variable $Q_r\equiv \frac{H}{\dot\barp}Q$ on super-horizon scales, coincides with the curvature perturbations at uniform density ($\zeta$). Thus, the constancy of $Q_r$ immediately implies the constancy of curvatures.
 The case is different in the oscillation period. In this case, the energy density (in our case the scalar field) cannot be used as a clock, and so curvature perturbations at a uniform density are not defined. In this case one has to re-interpret the constancy of $Q_r$ in terms of physical observables. This is precisely what will lead us to discover that curvature perturbations in Newtonian gauge may grow dangerously in Horndeski theories while $Q_r$ being still constant.\footnote{In \cite{Finelli:1998bu}, the Authors wondered whether, at linear level, $Q_r=const$ is still a solution in the points in which $\dot{\bar\phi} = 0$, given that the differential equation for $Q_r$ is proportional to $\dot{\bar\phi}^2 \dot Q_r=0$, in the strict $k=0$ case. Requiring $Q_r$ to be a continuous function, implies $Q_r=const$ at any time. On the contrary, if $Q_r$ was not continuous, it would imply a discontinuity in the Newtonian (Bardeen) potentials. However, this case is only possible in the presence of a localised (in time) source to the Einstein equations. Thus, this proves that $Q_r=const$ at any times, in $k=0$. The case $k\neq 0$ is more involved and we shall not discuss it here. Nevertheless, at least in GR, numerical results~\cite{parry} show that at linear level, the next to leading order in $k$ simply provides oscillations around the constant solution.
}
 
Coming back to the standard case, the relation between the Mukhanov-Sasaki variable and the Newtonian potential is then easily found by the use of the momentum constraint equation, the $(t,i)$ component of Einstein equations. This equation relates $\delta\phi$ to $\Phi$. Defining $w=\frac{p}{\rho}$, where the background energy density is $\rho=T^t{}_t=\frac{1}{2}\dot\phi+V$ and the background pressure is $p=\frac13T^i{}_i=\frac{1}{2}\dot\phi-V$, one then finds the first integral for the Newtonian potential on super-horizon scales

\be\label{standard}
\dot\Phi+\left(\frac{5}{2}+\frac{3}{2}w\right)H\Phi=\frac{3}{2}(1+w)H\zeta_c\ ,
\ee
where the constant $\zeta_c$ is fixed by the inflationary initial conditions and weakly depends on the wavelength of the perturbations that grew from sub- to super- horizon size during inflation. In particular, $\zeta_c$ is the constant part of the curvature perturbations in unitary gauge calculated during inflation, see Appendix~\ref{app-b}. Note that the constraint \eqref{standard} is well defined in $\dot{\bar\phi}=0$. Conversely, the would-be equation of motion for $\Phi$ is not.
 
The general solution of (\ref{standard}) is the homogeneous and the particular one. It is easy to see that the homogeneous solution is decaying in time. The particular solution depends on $w$. Assuming a positive potential, $w$ is a periodic function of roughly constant amplitude anywhere between $-1$ and $1$. In other words, the coherent oscillations of a scalar generate an average constant $w$ \cite{Mukhanov:2005sc}. In this case, the average solution of the Newtonian potential is a constant with amplitude matching the curvature perturbations generated during inflation by a factor of order $1$. Thus, the inflationary perturbations are retained unaltered in $\Psi$ until they re-enter in the horizon and are observed in the CMB. 

Let us end this section by noticing that the same equation \eqref{standard} can also be equivalently written as
\be\label{standard2}
\dot\Phi+\left(1+\epsilon\right)H\Phi=\epsilon H\zeta_c\ .
\ee

\section{Newtonian potential in Horndeski theories}

 The general Horndeski theory differs from the minimal case in a number of points. A striking difference is that Horndeski theories generate spacetime anisotropies, namely that $\Phi\neq \Psi$. Following \cite{Bellini:2014fua} one indeed has that, on super-horizon scales, the Newtonian potential $\Psi$ is related to the $\Phi$ potential by the following constraint
\begin{align}\label{psiphi}
\Psi=\Phi\left(1+\alpha_{\textrm{T}}\right)\left(1-\frac{\left(\alpha_{\textrm{M}}-\alpha_{\textrm{T}}\right)}{\epsilon+\left(\alpha_M-\alpha_T\right)}\right)
 -\frac{\left(\alpha_{\textrm{M}}-\alpha_{\textrm{T}}\right)}{H\left[\epsilon+\left(\alpha_M-\alpha_T\right)\right]}\dot{\Phi}\ ,
\end{align}
where the different parameters are re-listed in Appendix~\ref{app-a} for convenience of the reader. In \eqref{psiphi} we see that the Newtonian potential can become very large whenever the background approaches the pole $\epsilon+\left(\alpha_M-\alpha_T\right)\sim 0$. We will discuss this later on in a concrete example. In any case, we already see that this fact can potentially lead to the breaking of the linear regime in the intermediate region between inflation and the radiation dominated era. If this happens, the inflationary power spectrum cannot be transferred to the CMB.

Although now $\Phi$ and $\Psi$ are not the same, they are still related via the above constraint equation and thus, either one can be eliminated. If this was not the case, it would mean that the Horndeski theories would propagate more degrees of freedom.

Because of the derivative nature of the extra-coupling in Horndeski theories, these anisotropies are negligible during inflation and at low energies where the non-minimal coupling becomes small. On the contrary, the anisotropies can be large during the (intermediate) oscillating (reheating) stage. In any case, as in GR, once the $\Psi$ dependence of the perturbation equations is removed, it is possible to only consider $\Phi$, and the scalar perturbation $\delta\phi$. At this point, the traceless part of Einstein equation solves $\delta\phi$ in favour of background quantities and $\Phi, \dot\Phi$:
\be\label{condphi}
\delta\phi=-\frac{\left(1+\alpha_{\textrm{T}}\right)H\Phi+\dot\Phi}{\epsilon+\left(\alpha_{\rm M}-\alpha_{\rm T}\right)}\frac{\dot\barp}{H^2}\ .
\ee
Let us now consider the Mukhanov-Sasaki variable \eqref{conservation}. From there we can solve for $\dot\Phi$, plugging it into the momentum constraint we can solve for $\Phi$ (we remind the reader that after all there is only one scalar and we are treating it with $Q$). We are not showing the intermediate steps as they are cumbersome and not particularly enlightening. At this point, all variables depend on $Q$ and its derivatives. The Hamiltonian constraint again reduces to
\be\label{cons}
\partial_t\left(\frac{H}{\dot\barp}Q\right)=0\ ,
\ee
precisely as in GR. However, here, the functional form of $Q$ in terms of the Bardeen potential $\Phi$ is way different.  The fact that a super-horizon conserved variable {\it do} exist in this system should not be a surprise. As we saw earlier, the non-vanishing of $\Delta$ was solely due to a time-varying Planck mass $M_*$. Thus, as discussed in the introduction, we could already expect a conserved quantity in this system. Here, we explicitly found it to be $\frac{H}{\dot{\bar\phi}}Q$. The question is now, what this conserved quantity corresponds to, physically.

The conservation equation \eqref{cons} can be easily recast in terms of $\Phi$
\be\label{sol}
\dot\Phi+\left(1+\epsilon+\alpha_M\right)H\Phi=C H\left[\epsilon+\left(\alpha_M-\alpha_T\right)\right]= \zeta_c H\left[\epsilon+\left(\alpha_M-\alpha_T\right)\right]\ ,
\ee
where the constant $C$ depends on the initial conditions $\zeta_c$ settled during inflation (see Appendix~\ref{app-b}). In the canonical case in which $\alpha_i=0$, we recover the GR case, as expected. 

Although the general $\Psi$ equation \eqref{sol} closely resembles the minimal case, the qualitative behaviours of the Bardeen potentials are generically way different from the minimal case. Part of the solution of the equation \eqref{sol} is 
\be
\Phi\propto\exp\left[-\int \left(1+\epsilon+\alpha_{\rm M}\right)H dt\right]\ .
\ee
In GR, $\alpha_M=0$ and $\epsilon>0$ thus this mode decays. In the general Horndeski theory instead, one can easily have $\left(1+\epsilon+\alpha_M\right)<0$, transforming this mode into a growing mode. 

Similarly, the Newtonian potential $\Psi$ can grow and even worse go to large values when getting close to its pole, as already discussed.

The growing of the $\Phi$ and/or $\Psi$ potentials is potentially very dangerous. If $\Phi$ (or $\Psi$) reaches ${\cal O}(1)$, perturbation theory can no longer be used in any gauge and predictions of inflation are lost. On the other hand, requiring perturbation theory to be valid puts bounds on the amplitude of primordial power spectrum. In the next section, we will consider the concrete example of the new Higgs inflation of \cite{Germani:2010gm}. 

\section{A specific example: non-minimal kinetic coupling}

We will now discuss a model of inflation where the scalar field is non-minimally kinetically coupled to gravity with the unique interaction that does not introduce any new degree of freedom \cite{Germani:2010gm}
\be
{\cal L}_{\rm kin}=-\frac{1}{2}\left(g^{\alpha\beta}-\frac{G^{\alpha\beta}}{M^2}\right)\partial_\alpha\phi\partial_\beta\phi\ .
\ee
In the previous language that means:
\be
K=V\ ,\ G_3=-\frac{\phi}{2}\ ,\ G_4=\frac{M_p^2}{2}\ ,\ G_5=-\frac{\phi}{2 M^2}\ ,
\ee
where $V$ is the scalar field potential and all the other functions are taken to zero. Finally $M$ is a scale that parameterise the point in which the system becomes non-minimal.

This interaction was first considered in the so-called new Higgs inflation \cite{Germani:2010gm} and then in \cite{uv} for "natural" axionic inflationary scenario. The kinetic interaction of the Einstein tensor to gravity increases the gravitational friction during inflation by a factor $\propto \frac{3H^2}{M^2}\gg 1$, making any potential able to generate cosmic inflation\footnote{The proof of unitarity of this theory during inflation is given in \cite{Germani:2010gm} and in \cite{nico} by dimensional arguments. See also \cite{kunimitsu} for a quantum consistency of the large-field models in generalized G-inflation.}. The mechanism that does that was called gravitationally enhanced friction in \cite{gef} (see also \cite{rom} for a physical explanation) and the (slow) scalar so coupled to gravity was then called the "slotheon" in \cite{slo}. Scalar and tensor perturbations in unitary gauge were firstly considered in \cite{perturbations}. However, since we are interested in the oscillating phase where the unitary gauge is ill-defined due to switching the sign of  $\dot\phi$, we will work as before in the Newtonian gauge. 

Although the action is non-minimal, we may still recast this system in the Einstein equations form, i.e. $G_{\alpha\beta}=\frac{T^*_{\alpha\beta}}{M_p^2}$ where $T^*_{\alpha\beta}$ is a complicated energy-momentum tensor that depends on the scalar field and its up-to second order derivatives and on curvatures \cite{sus}. Finally, $M_p$ is the Planck mass.  

From the Einstein equations, it is easy to see that $w=-1-\frac{2\dot H}{3H^2}$.\footnote{Note that $w = \frac{p}{\rho}=\frac{p_*}{\rho_*}$.} The Friedmann equation is \cite{Germani:2010gm} 
\be
H^2=\frac{1}{3M_p^2}\left[\frac12\dot{\bar\phi}^2\left(1+\frac{9 H^2}{M^2}\right)+V\right]\ .
\ee
Taking the derivative of the Friedmann equation, we find that 
\be\label{dH/H2}
-\frac{\dot H}{H^2}=\frac{\dot M_*}{M_* H}+\ldots\ ,
\ee
where $\dot M_*\propto \ddot\barp$ and $\ldots$ stand for the ``standard" GR-like part. 

The decrease of $H$ in time implies that the gravitational enhanced friction becomes less and less efficient. In turn, $\dot M_*$ increases with time as the acceleration ($\ddot\barp$) grows with decreasing friction. At very low energies however, the system relaxes to GR and thus $\dot M_*\rightarrow 0$. Consequently $\epsilon$, and so $w$, grow while the non-minimal coupling is large to then to decay to the GR value. The maximum value for $\epsilon$ is thus expected around the point in which $3 H^2\sim M^2$. As we are going to see numerically later on, this interpretation is correct and in particular, $w$ can grow up to very large values.

In the high friction regime, where we have $H^2\gg M^2$ and, in the large $w$ limit \footnote{In reality $w$ is an oscillating function, however, in the region we are interested in, its average over a period is way larger than one, as we shall numerically see in the example we will discuss later on.} we have
\be\label{modified}
\dot\Phi+3wH\Phi=3wH\zeta_c\ .
\ee
By using the definition of $w$ one finds
\be
\Phi\simeq \zeta_c\left(1+c H^2\right)\ ,
\ee
where $c$ is an integration constant of dimensions of length square. 

This solution is valid so long as $w\gg 1$. At low energies, when $|w|\leq 1$, we know instead that the system relaxes to GR and $\Phi=b\zeta_c$, where $b\leq 1$ \cite{Mukhanov:2005sc} if $V$ is a renormalizable potential. Thus, since $H$ decreases in time (the universe expands) $c$ has to be negative, i.e., $\Phi$ grows in time until reaching the constant GR value. Finally we also note that during inflation, in this model, $\Phi=\frac{\epsilon}{3}\zeta_c$ which is different from the GR case in which $\Phi=\epsilon\zeta_c$ \cite{Mukhanov:2005sc}.

An explicit calculation shows that the Newtonian potential $\Psi$ also grows with $\frac{\dot M_*}{H M_*}$. The point is then whether or not the Bardeen's potentials $\Phi$ and $\Psi$, violate the linear regime by becoming of order one during the oscillation post-inflationary period.

\subsection{Numerics: the case of new Higgs inflation}

In what it follows, as a working example we numerically solve the background and perturbation equations for the case of the new Higgs inflationary scenario \cite{Germani:2010gm}:

In the new Higgs inflation of \cite{Germani:2010gm}, the Inflaton is the associated to the Higgs boson and the potential can be recast to be $V=\lambda\frac{\phi^4}{4}$ (for quantum effective potential see \cite{divita}, here we will only focus on the classical analysis). 

Using current collider data, we fix $\lambda=0.13$ \cite{lhc}. In \cite{perturbations,gef,nico} it was found that the slow-roll parameter during inflation $\epsilon_I=\frac{1}{3N+1}$, where $N$ is the number of e-foldings during inflation. For illustrative purposes, we will fix $N=60$. Once $\epsilon_I$ is given, all parameters of the system are only tight to the value of the amplitude of scalar perturbations \cite{nico}
\be
\frac{M}{M_p}&=&9\times 10^{-6}\left(\frac{\cal P}{2 \times 10^{-9}}\right)^{3/4}\frac{\epsilon_I^{5/4}}{\lambda^{1/4}}\ ,\cr
\frac{H_I}{M_p}&=&4\times 10^{-4}\left(\frac{\cal P}{2\times 10^{-9}}\right)^{1/2}\sqrt{\epsilon_I}\ .
\ee
Where the observed value of scalar temperature anisotropies $\cal P$ is $2\times 10^{-9}$ and $H_I$ is the Hubble scale during inflation. 

\begin{figure}
  \includegraphics[width=\linewidth]{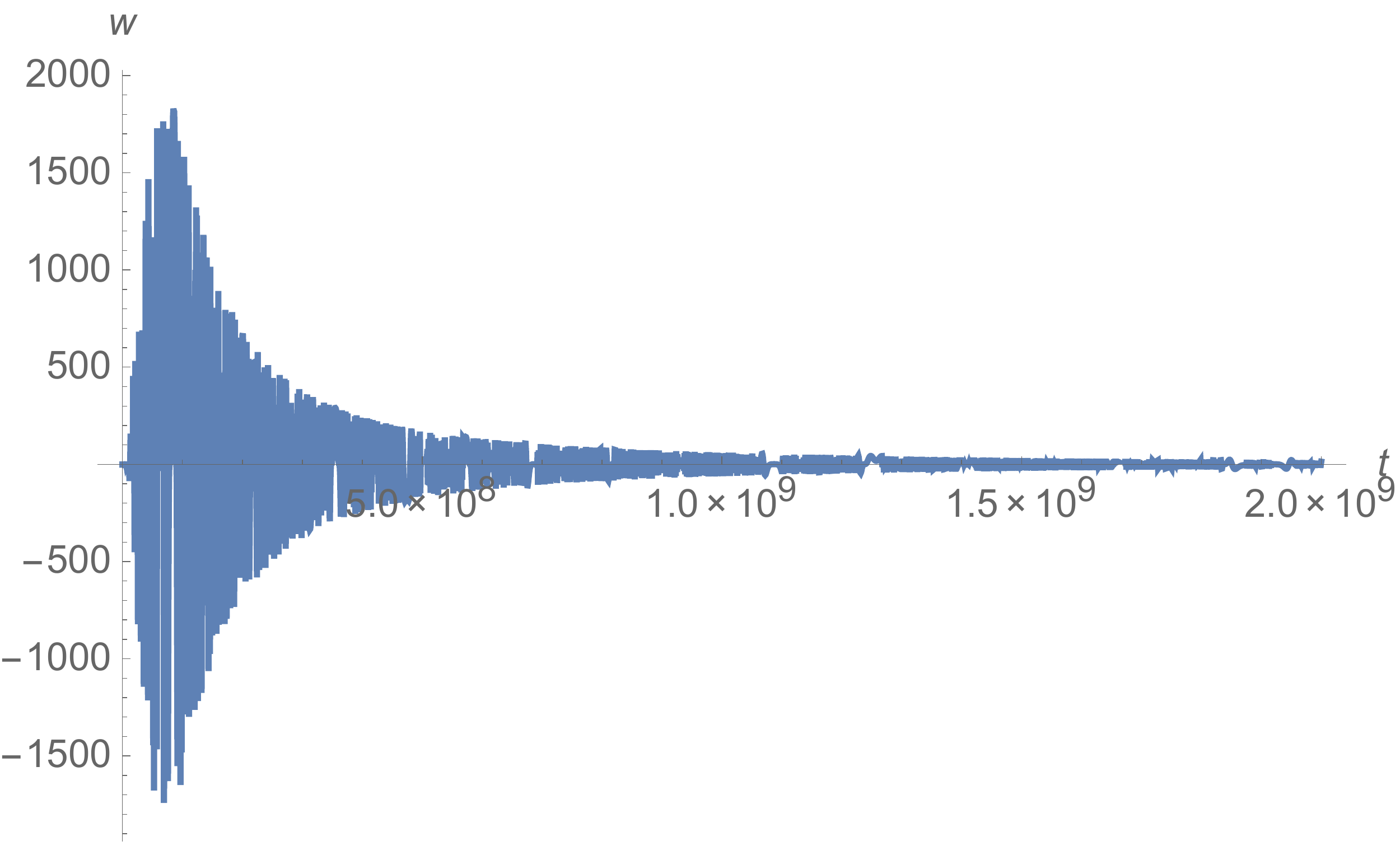}
  \caption{This is the effective $w$ in the new Higgs inflationary model where $N=60$ is fixed.}
  \label{fig1}
\end{figure}

\begin{figure}
  \includegraphics[width=\linewidth]{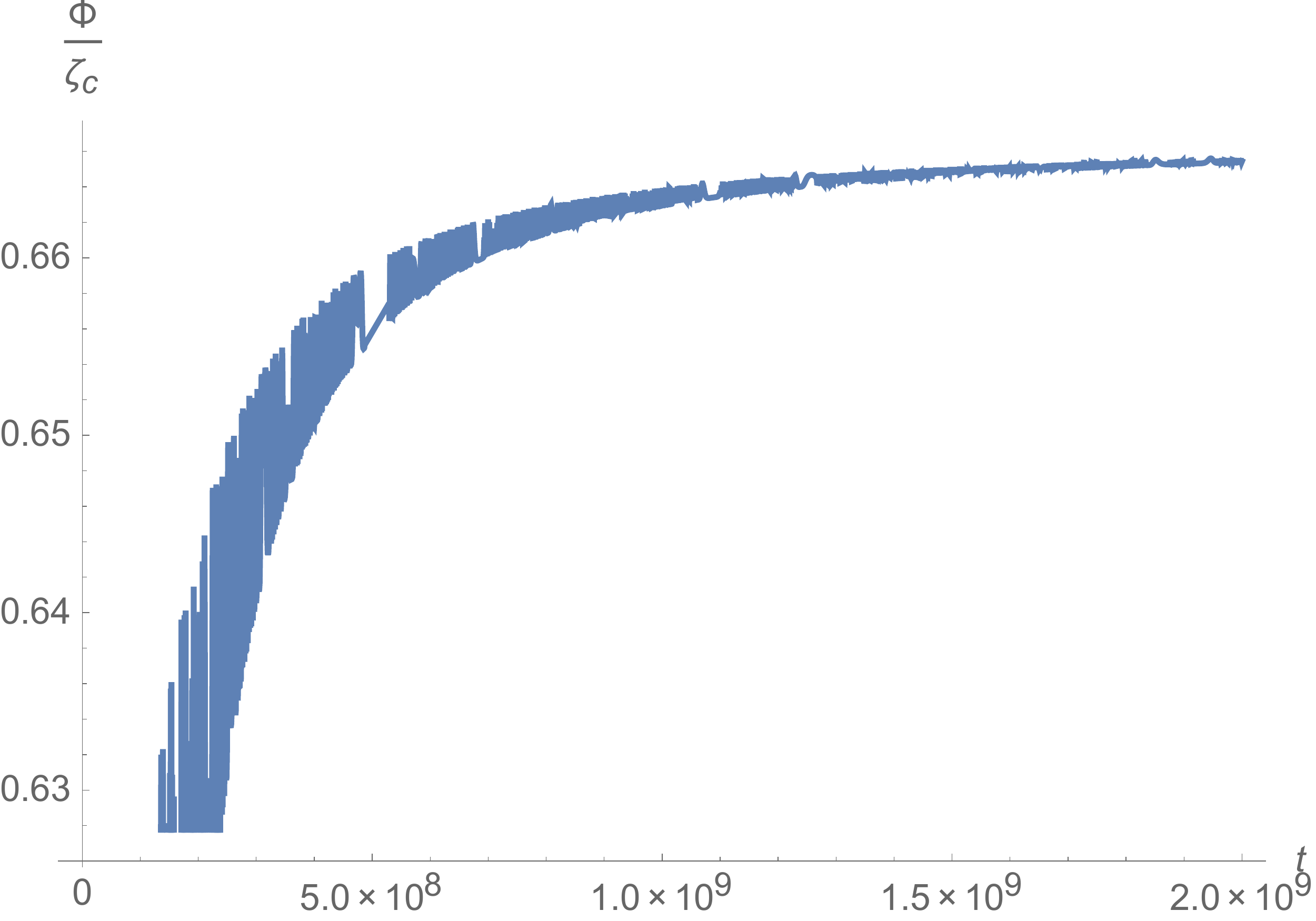}
  \caption{This is $\Phi$ in the new Higgs inflationary model where $N=60$ is fixed.}
  \label{fig2}
\end{figure}

\begin{figure}
  \includegraphics[width=\linewidth]{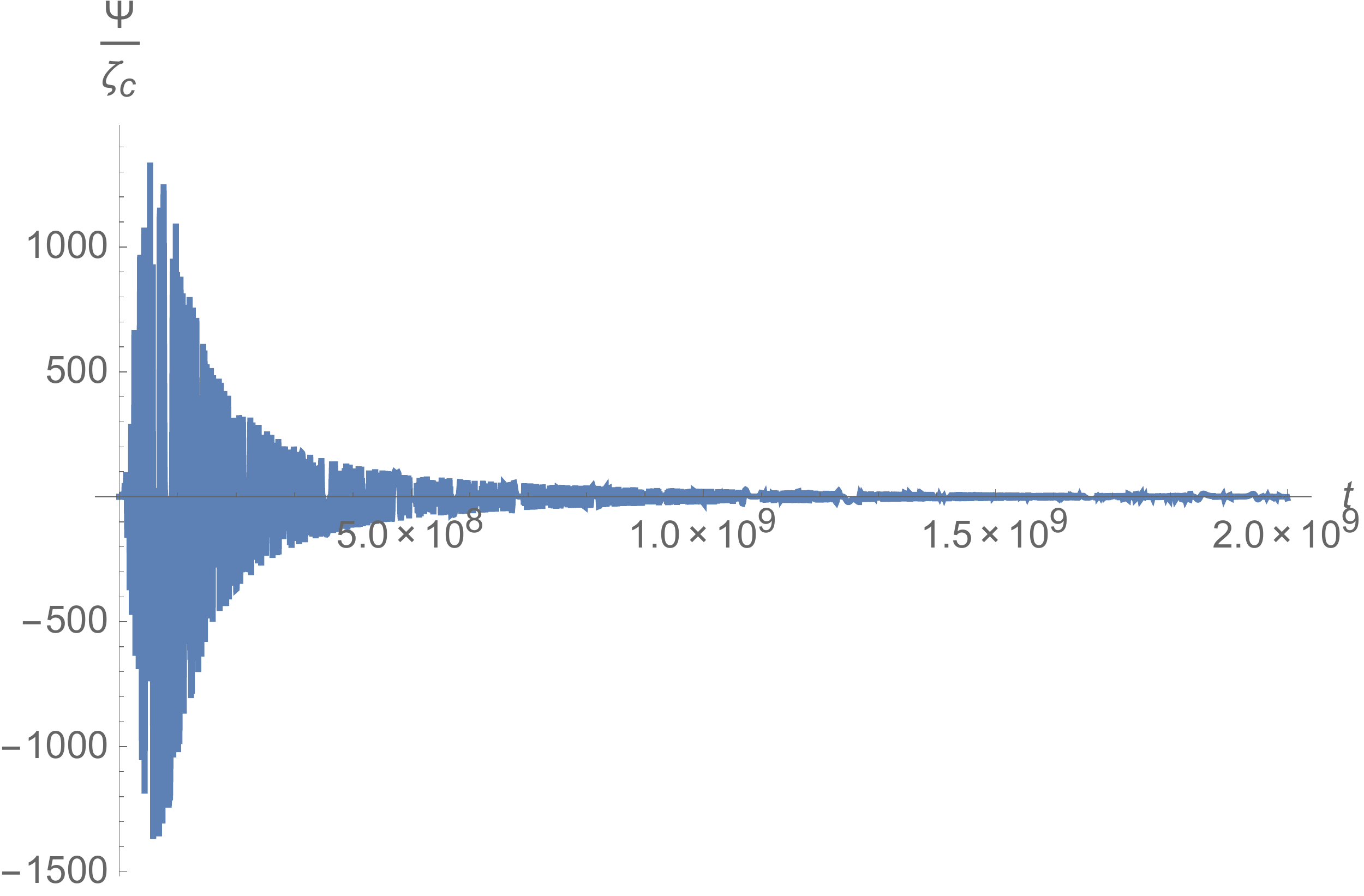}
  \caption{This is $\Psi$ in the new Higgs inflationary model where $N=60$ is fixed. We see the growing to large values of the Newtonian potential.}
  \label{fig3}
\end{figure}

\begin{figure}
  \includegraphics[width=\linewidth]{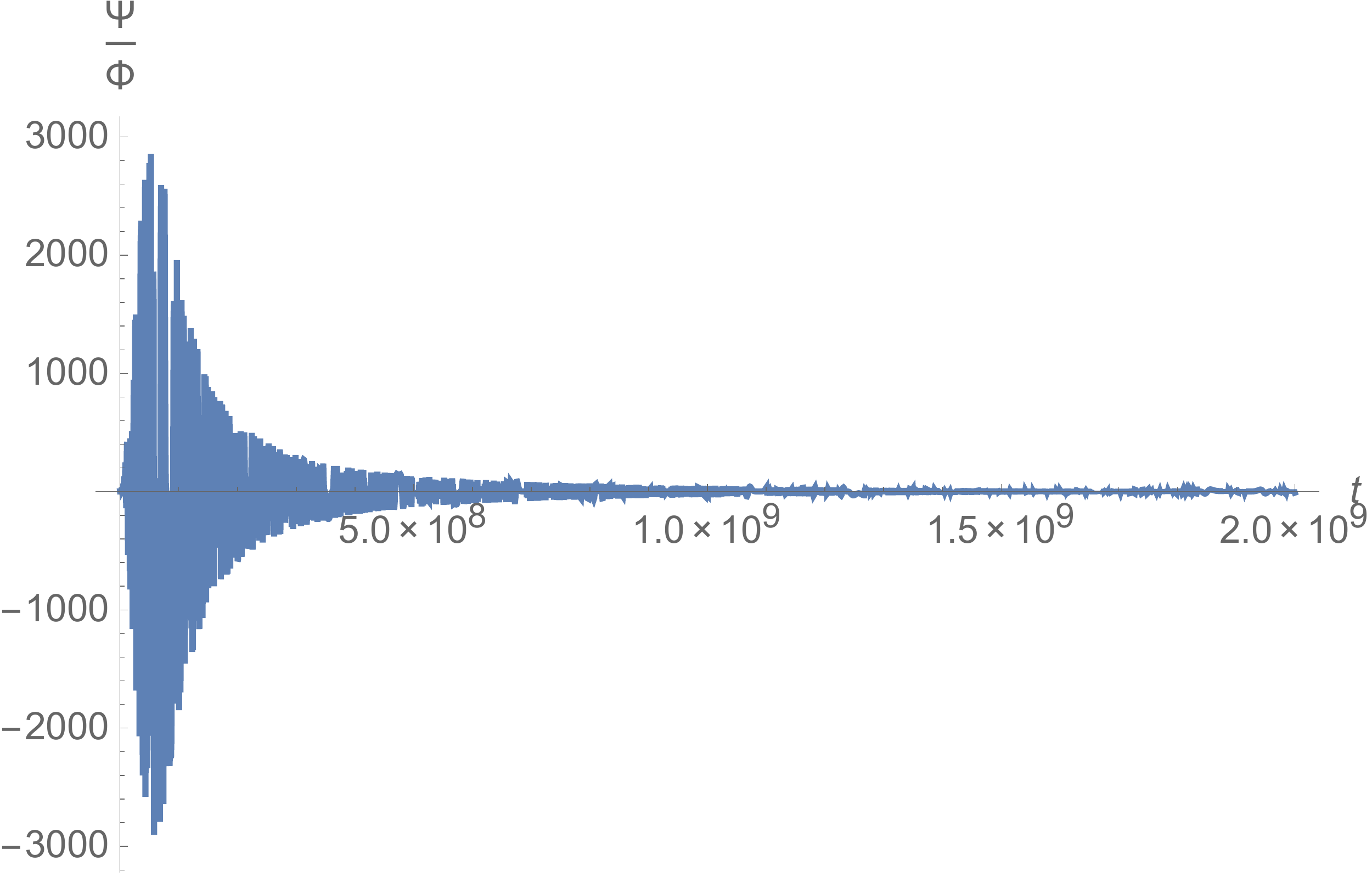}
  \caption{A measure of the anisotropy during the high-energy regime of the oscillating phase. Isotropy is for $\frac{\Psi}{\Phi}=1$.}
  \label{fig4}
\end{figure}

The behaviour of $w$ is shown in Fig.\ref{fig1}. From there it is clear that the new Higgs inflationary scenario undergoes a large $w$ regime after inflation. In the numerical units we are using, $M_p=1$ and the end of inflation happens roughly at $t_e=10^6$. The function $w$ grows after inflation until reaching a maximum. After that, $w$ decays to settle to the GR limit at low energy. As discussed, the growing of $w$ is related to the high friction regime, therefore the maximum happens when $H^2$ drops below $3M^2$ (as we also numerically checked).
In Fig.\ref{fig2} we instead plot the behaviour of $\Phi$ and in Fig.\ref{fig3} the Newtonian potential $\Psi$. As anticipated, $\Phi$ grows until settling to the GR value $\Phi=\frac{2}{3}\zeta_c$. On the contrary, the Newtonian potential $\Psi$ grows together with $w$ and then finally settles to the GR expected value $\Psi=\Phi=\frac{2}{3}\zeta_c$. The anisotropy is plotted in Fig.\ref{fig4}.

From the plots, it is clear that $\Psi\zeta_c^{-1}$ grows up to $\sim 10^3$. Thus, at the peak, the power spectrum would be of order $10^6\zeta_c^2$! In order to trust linear theory and avoid gravitational collapses into black holes, we must have $\Psi\ll 1/2$.  This puts a bound on the observed primordial power spectrum to be ${\cal P}=\zeta_c^2\ll 5\times 10^{-7}$. Fortunately, the observed power spectrum is of order $10^{-9}$ and thus no non-linear regime is reached before the radiation-dominated era. It is nevertheless interesting that the power spectrum is just right to avoid super-horizon black hole formations.

What we learned in this example is that generically, in Horndeski models, the post-inflationary physics can enormously amplify the Newtonian potential. Requiring the system to be kept in the linear regime poses severe limits on the validity of specific models of inflation that should be discussed case by case. For obvious reasons, we will not do it here.

\section{Conclusions}

Here we confirm that one can define a conserved super-horizon quantity, the rescaled Mukhanov-Sasaki variable ($Q_r$), throughout the Universe evolution. Whenever the scalar field velocity is sign-definite, this variable corresponds to the constant part of curvature perturbations at uniform density and thus is often confused with that. During the oscillating post-inflationary period however, uniform density surfaces cannot be defined. Thus one would need to physically interpret the constancy of the rescaled Mukhanov-Sasaki variable. In GR, the constancy of $Q_r$ immediately implies that the Newtonian potential is also almost constant (with a small jump from one effective equation of state to the other) so to assure the linear regime of perturbations throughout the Universe evolution.

In this paper we show that the Horndeski case is way different. The Newtonian potential can grow to potentially dangerously large values during the post-inflationary phase. This amplification can potentially violate the linearity of the system and thus must be discussed case by case. If this would happen indeed, the inflationary predictions would be completely lost. Since the growing of the Newtonian potential depends on the initial power spectrum, the linearity requirement may severely constrain the possible values of the observed power spectrum. Vice-versa, fixing the power spectrum to the observed one, the requirement of linearity could be a very powerful way to rule out non-minimally coupled models of inflation. 

However, if this does not happen, one can, as in the minimal case, encode the power spectrum of curvature perturbations at uniform density generated during inflation, in the CMB.\footnote{Of course we are assuming here that at low energies the system is the standard cosmological model.} This is basically due to the fact that the conserved variable $Q_r$ in terms of physical quantities matches the GR equivalent in the two extremes we are interested in: inflation and the radiation era. In fact, in this two regimes the system is approximately minimally coupled.

Whether or not this possible instability happens at smaller scales, will be precisely discussed elsewhere, however we will give here a few comments: 

The transition between super-and sub-``horizon" scales, as discussed in \cite{Bellini:2014fua}, is the braiding scale. In other words, whenever $\frac{k^2}{a^2}\gtrsim \frac{\beta_1}{\alpha_B^2}$ the system goes in the sub-horizon regime. 

At super-horizon scales, but finite wavelength of perturbations ($\lambda=k^{-1}< \infty$), the possible instability will persist by definition. 
At sub-horizon scales however, everything becomes trickier. Indeed, no conclusion can be drawn from the super-horizon behaviour. The reason is that small and large scale physics is separate. In other words, while the instability at super-horizon scales is roughly driven by anti-friction, a possible instability at sub-horizon scales can only be driven by some kind of sound speed. These two quantities are not necessarily connected (e.g. even in GR they are not). Thus, this analysis can only be done model by model. However, in this case, the equations for the $Q$ variable are not yet developed and we leave this for future work. In any case, what it is even worse, is that at small scales, the equations governing the Bardeen potentials are non-linear in $k^2$ \cite{Bellini:2014fua} (conversely to GR case) and thus, there is not even a precise definition of sound speed that leads our intuition on possible sub-horizon instabilities. 

Finally we have applied these discussion to the new Higgs inflationary scenario. There we indeed see a large amplification of the Newtonian potential, however still small enough to trust the linear regime.

\appendix
\section{The Bellini-Sawicki parameterization of Horndeski theories}\label{app-a}

All the equations that follow were derived in \cite{filippo2}, we will however use the compact notation of \cite{Bellini:2014fua}. The gravity equations on super-horizon scales can be so split into: 

The Hamiltonian constraint [$(tt)$ equation]:

\begin{align}
 & 3\left(2-\alpha_{\textrm{B}}\right)H\dot{\Phi}+\left(6-\alpha_{\textrm{K}}-6\alpha_{\textrm{B}}\right)H^{2}\Psi -\left(\alpha_{\textrm{K}}+3\alpha_{\textrm{B}}\right)H^{2}\dot{v}_{X}-\left[-3\dot{H}\alpha_{\textrm{B}}+6\dot{H}\right]Hv_{X}=0\,,\nonumber 
\end{align}
the momentum constraint [$(ti)$ equation]:

\begin{equation}
2\dot{\Phi}+\left(2-\alpha_{\textrm{B}}\right)H\Psi-\alpha_{\textrm{B}}H\dot{v}_{X}-2\dot{H}v_{X}=0\,,\label{eq:Momentum}
\end{equation}
the anisotropy constraint (spatial traceless part of the Einstein
equations)
\begin{equation}
\Psi-\left(1+\alpha_{\textrm{T}}\right)\Phi-\left(\alpha_{\textrm{M}}-\alpha_{\textrm{T}}\right)Hv_{X}=0\,,\label{eq:Aniso}
\end{equation}
and the pressure equation (spatial trace part of the Einstein equations)
\begin{align}
2\ddot{\Phi} & -\alpha_{\textrm{B}}H\ddot{v}_{X}+2\left(3+\alpha_{\textrm{M}}\right)H\dot{\Phi}+\left(2-\alpha_{\textrm{B}}\right)H\dot{\Psi}\\
 & +\left[H^{2}\left(2-\alpha_{\textrm{B}}\right)\left(3+\alpha_{\textrm{M}}\right)-\left(\alpha_{\textrm{B}}H\right)^{.}+2\dot{H}\right]\Psi\nonumber \\
 & -\left[2\dot{H}+\left(\alpha_{\textrm{B}}H\right)^{.}+H^{2}\alpha_{\textrm{B}}\left(3+\alpha_{\textrm{M}}\right)\right]\dot{v}_{X}\nonumber \\
 & -\left[2\ddot{H}+2\dot{H}H\left(3+\alpha_{\textrm{M}}\right)\right]v_{X}=0\,,\nonumber 
\end{align}
where $v_X\equiv-\frac{\delta\phi}{\dot\barp}$.

The following definitions were also introduced
\begin{align}
M_{*}^{2}\equiv & 2\left(G_{4}-2XG_{4X}+XG_{5\phi}-\dot\barp HXG_{5X}\right)\ ,\label{eq:planckmass}\\
HM_{*}^{2}\alpha_{\textrm{M}}\equiv & \frac{\mathrm{d}}{\mathrm{d}t}M_{*}^{2}\ ,\label{eq:omega1-1}\\
H^{2}M_{*}^{2}\alpha_{\textrm{K}}\equiv & 2X\left(K_{X}+2XK_{XX}-2G_{3\phi}-2XG_{3\phi X}\right)+\label{eq:omega2}\\
 & +12\dot\barp XH\left(G_{3X}+XG_{3XX}-3G_{4\phi X}-2XG_{4\phi XX}\right)+\nonumber \\
 & +12XH^{2}\left(G_{4X}+8XG_{4XX}+4X^{2}G_{4XXX}\right)-\nonumber \\
 & -12XH^{2}\left(G_{5\phi}+5XG_{5\phi X}+2X^{2}G_{5\phi XX}\right)+\nonumber \\
 & +4\dot\barp XH^{3}\left(3G_{5X}+7XG_{5XX}+2X^{2}G_{5XXX}\right)\ ,\nonumber \\
HM_{*}^{2}\alpha_{\textrm{B}}\equiv & 2\dot\barp\left(XG_{3X}-G_{4\phi}-2XG_{4\phi X}\right)+\label{eq:omega3}\\
 & +8XH\left(G_{4X}+2XG_{4XX}-G_{5\phi}-XG_{5\phi X}\right)+\nonumber \\
 & +2\dot\barp XH^{2}\left(3G_{5X}+2XG_{5XX}\right)\ ,\nonumber \\
M_{*}^{2}\alpha_{\textrm{T}}\equiv & 2X\left(2G_{4X}-2G_{5\phi}-\left(\ddot\barp-\dot\barp H\right)G_{5X}\right)\label{eq:omega4}\ .
\end{align}

\section{$C= \zeta_c$ during inflation}\label{app-b}
In the unitary gauge, the metric perturbation can be written as (note that ${\cal R}$ in \cite{kodama}, $\psi$ in \cite{nambu, naruko}, and $\zeta$ in \cite{gef, perturbations, kunimitsu, Maldacena} correspond to $-\zeta$ here)
\be
ds^2=-(N^2-N_iN^i)dt^2+2N_i dx^i dt+(1-2\zeta)a^2(t)d{\bm x}\cdot d{\bm x}\ ,
\ee
and the scalar field fluctuations are set to zero, i.e. $\delta\phi=0$. $N$ is a non-propagating scalar called the lapse and $N_i$ a non-propagating three-vector called the shift. We have ignored the tensor perturbations since scalar and tensor modes decouple at linear order.

Solving for the Hamiltonian and momentum constraints to the first order, and using their solutions, the quadratic action is given by~\cite{Kobayashi:2011nu} (see also \cite{tsujikawa} for the cubic action)
\be
S_{\zeta^2}=\int d^4x a^3\left[{\cal G}_S\dot\zeta^2-\frac{{\cal F}_S}{a^2}(\partial_i\zeta)^2 \right],
\ee
with the Bellini-Sawicki parameterization~\cite{Bellini:2014fua}
\be
{\cal G}_S &=& \frac{2M_*^2(\alpha_K +\frac32 \alpha_B^2)}{(2-\alpha_B)^2}\ , \\
{\cal F}_S &=& -\frac{2M_*^2\left[\dot{H}-\frac12H^2\alpha_B(1+\alpha_T)-H^2(\alpha_M-\alpha_T)\right]-H\dot\alpha_B}{H^2(2-\alpha_B)}\ ,
\ee
where the stability to the scalar modes requires ${\cal G}_S >0$ and ${\cal F}_S >0$.

In the unitary gauge, the two independent solutions on super-horizon scales are
\be\label{eq:zeta}
\zeta = \zeta_c + D\int\frac{dt}{a^3{\cal G}_S}\ ,
\ee
where $\zeta_c$ and $D$ are constant.
Since ${\cal G}_S\sim {\cal O}(\epsilon)M_p^2$ during inflation,\footnote{We consider inflation in an attractor phase. In a non-attractor phase~\cite{sasaki} $\epsilon$ could decay faster than $a^{-3}$, but we do not consider such a case.} as in the canonical single-field scenario~\cite{Maldacena}, the second term is obviously a decaying mode with $D \sim {\cal O}(k^2)$.\footnote{The unitary gauge is ill-defined when $\dot\phi=0$, and at these points the curvature perturbations can grow without violating causality. See \cite{naruko, gao} for fully non-linear conservation of the comoving curvature perturbation on super-horizon scales if the perturbation is adiabatic.}

In the Newtonian gauge, we have found the first integral for the rescaled Mukhanov-Sasaki variable that is conserved on super-horizon scales:
\be\label{c1}
\frac{H}{\dot{\bar\phi}}Q = \frac{H}{\dot{\bar\phi}}\delta\phi +\Phi = C\ .
\ee 
As done in the text, we can use the constraint equation \eqref{condphi} to relate $\delta\phi$ with $\Phi$ and its first derivative. By changing gauge to the unitary gauge, we finally find for $k\approx 0$
\be
\frac{H}{\dot{\bar\phi}}Q = C \approx \zeta= \zeta_c\ .
\ee

\acknowledgments

We would like to thank Tsutomu Kobayashi, Atsushi Naruko, Shinji Tsujikawa, and Filippo Vernizzi for correspondence.
CG would like to thank Emilio Bellini for many discussions and clarifications on Horndeski theories.  CG would like to thank Humboldt foundation for supporting the initial stages of this project. CG is supported by the Ramon y Cajal program and by the Unidad de Excelencia Mar\'{i}a de Maeztu Grant No.~MDM-2014-0369.
CG was also partially supported by the FPA2013-46570-C2-2-P grant.
NK would like to thank Gia Dvali and the European Research Council that partially supported this project. 
YW is supported by the JSPS Grant-in-Aid for Young Scientists (B) No.~16K17712.
YW was also supported in the initial stages of this project by the Grant-in-Aid for JSPS Fellows No.~269337.

\end{document}